\documentclass{article}

\usepackage{graphicx}
\usepackage{amsmath}
\usepackage{amsfonts}
\usepackage{amssymb}
\usepackage{natbib}
\bibpunct{(}{)}{;}{a}{}{,}

\newcommand{\comment}[1]{}

\begin{document}

\title{Does PIONEER measure local spacetime expansion?}

\author{Hans-J\"org Fahr and Mark Siewert\\
Argelander-Institut f. Astronomie, Abt. Astrophysik\\
{hfahr@astro.uni-bonn.de} \ \ \ {msiewert@astro.uni-bonn.de}}

\date{\today}

\maketitle

\maketitle

\begin{abstract}
There is a longstanding mystery connected with the
radiotracking of distant interplanetary spaceprobes like ULYSSES, Galileo and
especially the two NASA\ probes PIONEER 10 and 11. Comparing radiosignals
outgoing from the earth to the probe and ingoing again from the probes do show
anomalous frequency shifts which up to now have been explained as caused by
anomalous non-Newtonian decelerations of these probes recognizable at solar
distances beyond 5 AU.
In this paper we study cosmological conditions for the
transfer of radiosignals between the Earth and these distant probes.
Applying general relativity, we derive both the geodetic deceleration
as well as the cosmological redshift and compare the resulting
frequency shift with the observed effect.
We find
that anomalous decelerations do act on these probes which are of cosmological
nature, but these are, as expected from standard cosmology,
much too low to explain the observed effect. In contrast, the cosmological
redshift of radiophotons suffered during the
itinerary to the probe and back due to the local spacetime expansion
reveals a frequency shift which by its magnitude is
in surprisingly good agreement with the long registered phenomenon,
and thus explains the phenomenon well,
except for the sign of the effect. Problems of a local Hubble
expansion may give the reason for this.
\end{abstract}

\section{Introduction to the PIONEER\ phenomenon}
Besides other fundamental problems in present astrophysics and cosmology, for
instance connected with the nature of black holes, Quasars, Gamma ray
bursters, dark matter or dark energy, there exists
since about 20 years now the well recognized fundamental problem connected
with an anomalous deceleration towards the Sun registered at the motion of the
deep space probes like PIONEER-10 and 11 \citep[see][]{anderson98}.
Meanwhile these anomalous decelerations have also been recognized at the
spaceprobes Galileo and Ulysses \citep{anderson2002}. Nevertheless the
PIONEER\ spacecraft are especially appropriate for dynamical astronomy studies
due to the very accurate radiotracking operating for them. Due to their
spin-stabilization their acceleration estimates come down to the level of
$10^{-10} cm/s^{2}.$ The
VOYAGER\ spacecraft in this respect are much less suited for precise celestial
mechanics experiments as they perform too many attitude controle maneuvers
overwhelming all small external accelerations.

Since 1980, when PIONEER-10 moved at solar distances larger than 20 AU and the
Newtonian solar gravity pull dropped to levels of $\alpha_{s}\leq5\cdot
10^{-8}cm/s^{2}$, the JPL\ orbit determination program (ODP) found unmodelled
accelerations with a systematic residual level of $\alpha_{a}\simeq
(8.74\pm1.33)\cdot10^{-8}cm/s^{2}$ directed towards the sun. Interestingly
enough the level of these residual decelerations, besides some 10 percent
fluctuations, remained constant for all the ongoing PIONEER\ itinerary to
larger distances, i.e. seemed to prove as being independent on solar distance,
orientation and time.
A large number of proposals how these anomalous decelerations could
perhaps be explained have meanwhile been proposed (see \cite{anderson2002}
or \cite{dittus2006}). Amongst them one finds friction forces with
interplanetary dust grains, asymmetric thermal emission from the probe, an
accelerated motion of the whole solar system in the direction normal to the
ecliptic, MOND\ effects or dark matter gravity contributions have been
discussed, but none of these proposed explanations up to now could fit the
observed magnitude and the distance-independence of the anomalous
deceleration. For this reason more recently also cosmological causes for the
existing anomalous deceleration have been suspected, all the more because the
value found for the anomalous deceleration nicely is represented by the
cosmological quantity $\alpha_{a}\simeq cH_{0}$, where $H_{0}$ denotes the
present-day Hubble constant of about the order of 70 km/s/Mpc.

\section{The local equation of geodetic motion}

Spoken in terms of general relativity, the solar system is embedded
into a local spacetime metric which is not of purely cosmological nature,
but locally has an imprint from gravitational binding forces of
gravitationally bound masses of our host galaxy, i.e.the milky way.
It should be noted that there is some doubt that
local cosmological expansion does even exist below the scale of
galaxy clusters \citep{bk-misner}.
However, there are results by other authors who believe differently
\citep[see][and references therein, for a recent review]{bonnor2000},
and the final answer to this question is still to be given.
For this reason we now try to estimate what kind of forces (and other
effects) would result from such a contribution.

The first attempt to describe gravitationally bound masses embedded in the
Robertson-Walker metric of an expanding universe goes back to
\cite{es45}. As shown there the connection of an outer Schwarzschild
metric of a central mass $M$ to an outer cosmic Robertson-Walker
metric seems possible at a critical distance $r_{ES}$ from the central
mass, which is called the Einstein-Straus vacuole. This critical radius
is given by:
\begin{equation}
r_{ES} = \left( \frac{3M}{4\pi \rho_0} \right)^{1/3}.
\end{equation}

It turns out that, since in an expanding universe, the mass density
$\rho_0$ is decreasing with time, the radius of the
ES-vacuole increases with time according to
\begin{equation}
\frac{\dot{r}_{ES}}{r_{ES}} = \frac{\dot{R}}{R} = H_0.
\end{equation}
The main problem connected to the ES-vacuole is that, in the present
time, it is very large, i.e. $r_{ES}(1 M_\odot) \simeq 100 pc$
and $r_{ES}(1 M_{Gal} = 10^{11} M_\odot) \simeq 0.5 Mpc$.
This means that, if ES-vacuoles would really surround
bound systems, such as galaxies, then essential fractions of cosmic space
would be described by the static Schwarzschild metric within these
vacuoles instead of the Robertson-walker one. This relation
has also been pointed out by \cite{cg06}. For the
well-established cosmological photon redshift, this means that
this quantity would be no longer related to the cosmological distance,
invalidating one of the most fundamental results from cosmology.

The problem of the actually prevailing cosmic metrics within
gravitationally bound systems is still highly controversial at present,
as demonstrated in a recent paper by \cite{cg06}.
These authors also demonstrate, using clear astrophysical arguments, that
the answer given by \cite{es45} and successor papers is not
able to adequately solve this problem. They argue in their paper
that in an expanding Robertson-Walker universe, to keep a test
mass $m$ at a timelike distance $r$ from the center of the solar gravity,
a cosmological acceleration has to operate which is given by:
\begin{equation}
|\ddot{r}|_{CA} = \left( \frac{\ddot{R_0}}{R_0} \right) r = - q_0 H_0^2 r,
\end{equation}
where $R_0$, $q_0$ and $H_0$ denote the present scale of the universe,
the acceleration parameter and the Hubble constant. The classical
Newtonian gravity force hence only includes the residual acceleration
and leads to the following equations of motion:
\begin{equation}
m (\ddot{r} - |\ddot{r}|_{CA}) = - \frac{C}{r^2} + \frac{L^2}{r^3}
\end{equation}
and:
\begin{equation}
L = r^2 \phi
\end{equation}
where $C$ contains the central mass of the sun, $L$ is the angular momentum
of the orbiting object, and the azimuth has been denoted with $\phi$.
These modified equations can also be applied to an object like the PIONEER
spacecraft, but it clearly turns out that the cosmological acceleration
is unable to explain the anomaleous frequency shifts of the
PIONEER signals.

We consider the PIONEER spacecraft as an object carrying out only a
geodetic motion in the local spacetime metrics of the surrounding universe,
with a time-dependent Robertson-Walker-like metric with a
local scale parameter $L$ (LRW-metric) and the associated
Christoffel symbols, which in turn are a function of first
partial derivatives of the metrical tensor $g_{ik}$,
\begin{equation}
\Gamma^i_{jk} = \frac{1}{2} g^{il}
 (\partial_j g_{lk} + \partial_k g_{lj} - \partial_l g_{jk} ).
\end{equation}
The general-relativistic generalisation of the simple Newtonian
force-law then is
\begin{equation}
\frac{d^{2}x_{\alpha}}{ds^{2}}+\Gamma_{\mu\nu}^{\alpha}\frac{dx_{\mu}}%
{ds}\frac{dx_{\nu}}{ds}=f^{\alpha},
\end{equation}
with the spacetime coordinates $x_{\alpha}=\left\{  x_{0},x_{1},x_{2},x_{3}\right\}
=\left\{ct,r,\vartheta,\phi\right\}$ and the world line element
$ds=c\ d\tau =c\gamma(v) dt$. The expression $f^\alpha$ is the
four-force/mass, representing the gravitational pull of the sun, and
other external forces, while $\gamma(v)=(\sqrt{1-v^2/c^2})^{-1}$
is the well known relativistic Lorentz factor. The coordinate system has been
selected in a way that the center ($r=0$) is located in the sun
(i.\ e.\ approximately at the earth).

Assuming now that the four-force/mass does systematically vanish
at larger distances from the sun, one is left with the
geodetic equation for freely moving particles
\begin{equation}
\frac{d^{2}x_{\alpha}}{ds^{2}}+\Gamma_{\mu\nu}^{\alpha}\frac{dx_{\mu}}%
{ds}\frac{dx_{\nu}}{ds}=0.
\label{eq-geodetic}
\end{equation}

For a purely radial motion, i.e. an object only changing the increment of $r$
at its motion, with $d\vartheta,d\phi=0$, eqn. \ref{eq-geodetic}
reduces to
\begin{equation}
\begin{split}
& \frac{d^{2}r}{ds^{2}}+\Gamma_{\mu\nu}^{r}\frac{dx_{\mu}}{ds}\frac{dx_{\nu}%
}{ds}
\\
= & \frac{d^{2}r}{ds^{2}} + \left\{
 \Gamma_{0\nu}^{r}\frac{dx_{0}}{ds}+\Gamma_{r\nu}^{r}\frac{dr}
{ds}\right\}  \frac{dx_{\nu}}{ds}
\\
& +\left\{  \Gamma_{\nu0}^{r}\frac{dx_{0}}
{ds}+\Gamma_{\nu r}^{r}\frac{dr}{ds}\right\}  \frac{dx_{\nu}}{ds}
\\
=& 0.
\end{split}
\end{equation}
After a quick sorting this expression further reduces to
\begin{equation}
\begin{split}
\frac{d^{2}r}{ds^{2}} =& - 2 \left\{
   \Gamma_{00}^{r}(\frac{dx_{0}}{ds})^{2}
  +\Gamma_{r0}^{r}\frac{dr}{ds}\frac{dx_{0}}{ds}
\right.
\\
& \left.
  +\Gamma_{0r}^{r}\frac{dr}{ds}\frac{dx_{0}}{ds}
  +\Gamma_{rr}^{r}(\frac{dr}{ds})^{2}
\right\}.
\end{split}
\end{equation}

We now consider a local Robertson-Walker-type metric
(LRW metric), for which the coefficients are given by
\begin{align}
g_{tt} & =1 & g_{rr} & =-\frac{L^{2}(t)}{1-kr^{2}}
\nonumber
\\
 g_{\vartheta\vartheta} & =-L^{2}(t)r^{2} &
g_{\phi\phi} & =-L^{2}(t)r^{2}\sin^{2}\vartheta
\end{align}
\begin{align}
\Gamma_{rr}^{r} & = \frac{rk}{1-kr^{2}} & \Gamma_{00}^{r}&=0
\nonumber
\\
\Gamma_{0r}^{r} & =\frac{1}{c}\frac{\dot{L}}{L} & \Gamma_{r0}^{r} &=0,
\end{align}
where $k$ is the curvature parameter and $L(t)$ is the local
scale factor.

Thus one finally obtains the following differential equation in coordinates
$r$ and $s$
\begin{equation}
\frac{d^{2}r}{ds^{2}}{=-}\frac{2}{c}\frac{\dot{L}}{L}\frac{dr}{ds}%
\frac{dx_{0}}{ds}{-}\frac{2rk}{1-kr^{2}}{(}\frac{dr}%
{ds}{)}^{2}.
\end{equation}
For subrelativistic object velocities, i.e.
$\gamma(v)=1$, this expression transforms into
\begin{equation}
\frac{d^{2}r}{dt^{2}} = -\frac{2 \dot{L}}{L}\frac{dr}{dt}{-}%
\frac{2rk}{1-kr^{2}}{(}\frac{dr}{dt}{)}^{2}.
\end{equation}
Assuming that the universe is flat ($k=0$), this expression
further reduces to
\begin{equation}
\frac{d^{2}r}{dt^{2}} = -\frac{2 \dot{L}}{L}\frac{dr}{dt}.
\label{eq-d2sdt2}
\end{equation}

For a purely radial motion, the metric line element is given
by the expression
\begin{equation}
ds_{r}=\sqrt{-g_{rr}}dr=L\frac{dr}{\sqrt{1-kr^{2}}}.
\end{equation}
Using again $k=0$, this relation simplifies to
\begin{equation}
s_{r}=Lr,
\end{equation}
and the radial velocity becomes
\begin{equation}
v_{r}=\frac{ds_{r}}{dt}=\dot{L}r+L\frac{dr}{dt},
\label{eq-vradial}
\end{equation}
which can be rearranged to
\begin{equation}
\frac{dr}{dt}=\frac{1}{L}(v_{r}-\dot{L}r).
\label{eq-dsdt}
\end{equation}

Differentiating eqn. \ref{eq-vradial} leads to an equation for
the radial acceleration $b_r$,
\begin{equation}
b_{r}=\frac{d^{2}s_{r}}{dt^{2}}=\ddot{L}r+2\dot{L}\frac{dr}{dt}+L\frac{d^{2}.
r}{dt^{2}}%
\label{eq-br1}
\end{equation}
Applying eqns. \ref{eq-d2sdt2} and \ref{eq-dsdt}, it is possible to
eliminate all nonmetric quantities by metric ones and the
differential equation \ref{eq-br1} turns into the metrical equation
\begin{equation}
{b}_{r} = s_{r} \frac{\ddot{L}}{L}.
\label{eq-br2}
\end{equation}
This expression is nonzero if we assume that the universe
is not only expanding, but accelerating, which has been verified
by recent observations of distant supernovae \cite{perlmutter2003}
as well as CMB radiation by the WMAP experiment \citep{wmap}.
On the other hand, this expression depends on the distance of the
spacecraft from the sun, which contradicts the distance-independence
of the observed PIONEER effect \cite[see][]{anderson2002}.

If we assume that cosmological expansion is an upper limit
to what may happen on smaller scales, i.\ e.\ $\ddot{L}=\ddot{R}$,
then, using
\begin{equation}
q_0 = - \frac{\ddot{R}R}{\dot{R}^2} \simeq \frac{1}{2},
\end{equation}
we obtain
\begin{align}
b_r &< - s_r q_0 H_0^2.
\end{align}
Interestingly, this result is almost identical to the result
from \cite{cg06}, where the relation $\ddot{a}/a \simeq -q_0 H_0^2$
was found.
From these two equations it follows that, in order to reproduce
$b_r \simeq H_0 c$, and using typical values of $s_r = o(10 AU)$
and $q_0 = o(1)$, we need
\begin{equation}
s_r = o(10^{26}) m \simeq o(10^{15}) AU.
\end{equation}
Thus, at the distances covered by the PIONEER spacecrafts, the
resulting force is much too weak to be observable.

To summarize our results, if cosmological expansion is
present on the length scale of the solar system, its effect
in terms of anomalous acceleration would
definitely not be able to explain the observed blueshift in
the PIONEER signal.

\section{The cosmological redshift of radiophotons}

Another, considerably different possibility to explain the
anomalous PIONEER acceleration, is to assume that the observed
effect is a ``fake'' acceleration, which in truth follows from
some fundamental misunderstanding about the underlying physics
of radiowave propagation in space.
While a considerable amount of work has been put into tracking
down additional gravitational sources, there have only been a few
publications investigating this other possibility in the past years.
\cite{rossg99} and \cite{ros02} have investigated the possibility
of a systematical error in the measurement of cosmological
distances and times, obtaining
a result with the correct order of magnitude. However, \cite{cg06}
claim that this estimate is wrong, and that the result
should be reduced by an order of $(v/c)^3$.

We now investigate a different possibility related to cosmological
expansion, namely the redshift suffered by
massless particles (such as radiophotons) freely propagating through
an expanding space.
According to JPL (ODP), the registered phenomenon of the PIONEER anomaly by
formula is represented in the form \citep{anderson2002}
\begin{equation}
\Delta\nu = \nu_0 - \nu_1
= - \nu_{0}\frac{2\textsl{a}_{PIO}t_{i}}{c},
\label{eq-anderson}
\end{equation}
where the time is normalized in a way that $2t_{i}$
denotes the passage time of the electromagnetic radio signal on
its way from the Earth to the spaceprobe and back to the receiver on
earth, and $a_{PIO}$ denotes the expected so-claimed residual acceleration
in the PIONEER\ motion, normalized in a way that $a_{PIO} > 0$ corresponds
to an acceleration towards the sun. The subscript $1$ denotes the quantities
at $t=2t_i$, when the returning photon is observed.
It should be noted that the above frequency shift is normalised in
a different way than usual, where a
redshift results in $\Delta\nu > 0$ and a blueshift in $\Delta\nu < 0$
\citep[see][ref. 38, the ''usual'' definition results in an additional negative
sign]{anderson2002}.

The cosmological wavelength-redshift relation
for a photon in a local Robertson-Walker-like spacetime (LRW-metric) is
\begin{equation}
\frac{\lambda_1}{\lambda_{0}}=\frac{\nu_{0}}{\nu_1}=\frac{L_1}{L_{0}},
\end{equation}
where the subscript $0$ denotes the respective value of the emitted
photon, while the subscript $1$ denotes the observed, redshifted photon.
Here we have called the initial scale factor $L(t=0)=L_0$
and the scale factor at $t=2t_i$ is called $L_1$. For this scale
factor and for short distances it is possible to expand
$L_1 = L_0 + \dot{L}_0 \cdot 2t_i$ and we obtain
\begin{equation}
\Delta\nu= \nu_{0}-\nu_1 =\nu_{0}\left(  1-\frac{L_{0}}
{L_1}\right)  =\nu_{0}\left(  1-\frac{L_{0}}{L_{0}+\dot{L}_{0}2t_{i}}\right).
\end{equation}
Assuming that $(\dot{L}/L) \cdot 2t_i \ll 1$,
this expression further simplifies to
\begin{equation}
\Delta\nu=\nu_{0}\left(  1-\frac{1}{1+\frac{\dot{L}_{0}}{L_{0}}2t_{i}}\right)
\simeq\nu_{0}\left( 1-1+\frac{\dot{L}_0}{L_0} 2t_{i}\right).
\end{equation}

If we compare this expression with eqn. \ref{eq-anderson}, we obtain
\begin{equation}
\Delta\nu = - \nu_{0}
\frac{\textsl{a}_{PIO}t_{i}}{c}= \nu_{0} \frac{\dot{L}_0}{L_0} t_{i},
\end{equation}
and finally
\begin{equation}
{a}_{PIO}=- \frac{\dot{L}}{L} c.
\label{eq-apio}
\end{equation}
For a full cosmological expansion, we obtain
\begin{equation}
{a}_{PIO}=- H_0 c.
\end{equation}
Except for the sign, this result is in remarkably
good agreement with the observed anomalous acceleration term,
$a_{PIO} \simeq H_0 c$.

Ignoring the wrong sign, we obtain numerically
for $a_{PIO} = (8.74 \pm 1.33) \cdot 10^{-10} m/\sec^2$
\begin{equation}
\begin{split}
H_0 &= |a_{PIO}| = (2.91 \pm 0.44) \cdot 10^{-18} s
\\
 &= (89.8 \pm 13.5) \frac{km}{s\ Mpc},
\end{split}
\label{eq-missingx}
\end{equation}
which is only marginally different from the observed value of the Hubble constant,
namely $H_0 = (70+2.4-3.2) km/(s Mpc)$.

\section{The local spacetime expansion}
\label{sec-local}

We consider the metric conditions of the space environment of our solar system
and of the milky way as it may be described by a local metrical perturbation
of the general cosmological spacetime metrics (i.e. the RW-metrics) due to the
enhanced matter density in the local cosmic environment compared to the
large-scale average of cosmic matter density. \cite{bonnor57} has described this
situation by introduction of a function called the local density contrast
$\delta$ and given by:%
\begin{equation}
\delta(L,t)=\frac{\rho(L,t)-\left\langle \rho(t)\right\rangle }{\left\langle
\rho(t)\right\rangle }=\frac{\Delta\rho(L,t)}{\left\langle \rho
(t)\right\rangle },
\end{equation}
where $\rho(L,t)$ and $\left\langle \rho(t)\right\rangle $ denote the average
density over a scale $L$ and the large-scale average
(i.\ e.\ $L \rightarrow L_\infty \simeq \infty$) of the cosmic matter
density, respectively.  For the growth of this function Bonnor has derived the
following differential equation:%
\begin{equation}
\ddot{\delta}+2\frac{\dot{R}}{R}\dot{\delta}-4\pi G\left\langle \rho
(t)\right\rangle \delta=0.
\end{equation}
The solution of this differential equation for an Einstein-DeSitter type
universe with $\Lambda=0$ and $k=0$ is given by (see, e.\ g.\ \cite{goenner97}
or \cite{silkbouwens2001})
\begin{equation}
\delta=\delta_{0}\left(  \frac{t}{t_{0}}\right)  ^{2/3},
\label{eq-deltat}
\end{equation}
where $\delta_{0}$ is the density contrast at some reference time $t_{0}$.

Considering galaxies with a typical scale $L$ and typical mass $M_{gal}$ one
can express the density contrast on the galactic scale by:%
\begin{equation}
\delta(L(t))=\frac{\frac{M_{gal}}{L^{3}}}{\frac{M_{U}}{R^{3}}}-1=\left(
\frac{M_{gal}}{M_{U}}\right)  \left(  \frac{R}{L}\right)  ^{3}-1,
\end{equation}
with the expression
\begin{equation}
\left\langle \rho(t)\right\rangle =
M_{U}/R(t)^{3}.
\end{equation}
From that relation we derive
\begin{equation}
\left(  \frac{M_{gal}}{M_{U}}\right)  \left(  \frac{R}{L}\right)
^{3}-1=\delta_{0}\left(  \frac{t}{t_{0}}\right)  ^{2/3}.
\end{equation}

Since the present day density contrast has grown to very large values
of the order of $\delta\simeq10^{6}$ we thus can find:%
\begin{equation}
L=R \left(  \frac{M_{gal}}{M_{U}}\right)  ^{1/3}\delta_{0}%
^{-1/3}(\frac{t_{0}}{t})^{2/9}=R \Gamma(\frac{t_{0}}{t})^{2/9},
\label{eq-loverr}
\end{equation}
with $\Gamma=\left(  \frac{M_{gal}}{M_{U}}\right)  ^{1/3}\delta_{0}^{-1/3}$.
By differentiating eqn. \ref{eq-loverr} we obtain
\begin{equation}
\dot{L} = \dot{R} - \frac{2}{9} \frac{L}{t},
\end{equation}
or
\begin{equation}
\frac{\dot{L}}{L} = H_0 \left( 1 - \frac{2}{9} \frac{T_{0}}{t} \right),
\end{equation}
where $T_0$ is the Hubble age of the universe and $t$ the time since
which the initial density contrast has been evolving, say, the
matter recombination era when $\delta_0$ was of the order of
$\delta_0 \simeq 10^{-5}$ \citep[see][]{wmap}. For a linearly
expanding universe we then get $t\simeq T_0$ and
\begin{equation}
\frac{\dot{L}}{L} = H_0 \left( 1 - \frac{2}{9} \right) = \frac{7}{9} H_0.
\end{equation}

Applying this expression to eqn. \ref{eq-apio} leads to
\begin{equation}
a_{PIO} = - \frac{7}{9} H_0 c.
\end{equation}
This means that the predicted redshift is no longer as red as expected
on a cosmological length scale, but it still spots the wrong sign.

Can we use this mechanism to obtain the observed
blueshift? This would require
\begin{equation}
\frac{7}{9} \frac{T_0}{T_r} - 1 \simeq 1,
\end{equation}
which in turn leads to a required true recombination
age of the universe ($T_r$) of
\begin{equation}
T_r \simeq \frac{7}{18} T_0,
\end{equation}
which isn't even half the (Hubble) age of the universe.

Comparing this
with the current experimental result that the universe seems
to be uncurved ($k=0$), and undergoing an accelerating expansion, the
only RW-like cosmological model which qualitatively fits these observations
is the variant using $\Lambda>0$, which
predicts an \emph{older} Universe than infered from the
Hubble constant \cite[see, e.g.][]{bk-inverno}.
From this we conclude that even after correcting
our results from the last section for local perturbations
by the galaxy in a (linear) Einstein-DeSitter type universe
our ansatz still can't explain the PIONEER anomaly.

On the other hand, the above derivation is based on the theory of
linear growth of density contrasts. Starting, however, from initial levels of
the order of $\delta_0 \simeq 10^{-5}$ at the recombination era,
density contrasts meanwhile have grown up to to strongly nonlinear
levels of the order of $\delta\simeq 10^6$ in the present universe. Thus
linear perturbation theories as they were applied in the derivations above
lead to misleading results. In case of strongly nonlinear growth
the local scale $L$ may, in fact, even still now undergo a local
cosmological blueshift instead of the redshift of freely propagating photons
in the milky way environment. Since analytical results
on the nonlinear growth of density contrasts do not exist yet, we
instead try an alternative way to describe the rate of a
local cosmological expansion. At the recombination era,
with a cosmological redshift of $z_r = 1000$, the cosmic density
was given by $\rho_r = 10^9 \rho_0$, where $\rho_0$ denotes
the present-day cosmic density. On the other hand, the present-day density
contrast on the galactic scale is given by
\begin{equation}
\delta_0(L_{gal}) = \frac{\rho_{gal} - \rho_0}{\rho_0} \simeq 10^6.
\end{equation}
This indicates that the mass density on the galactic scale has only decreased
by three orders of magnitude since the recombination era, clearly pointing
to the fact that the cosmic volume forming a present-day galaxy has expanded
differentially with respect to the rest of the universe, roughly given by
the expansion law
\begin{equation}
L_{gal,0} = L_{gal,r} \left( \frac{R_0}{R_r} \right)^\alpha,
\end{equation}
with $\alpha \simeq 1/2$. From this relation it is possible to derive the
expansion law
\begin{equation}
\frac{\dot{L}_{gal,0}}{L_{gal,0}} = \frac{1}{2} \frac{\dot{R}_0}{R_0}.
\end{equation}
On the basis of this relation we estimate the resulting frequency shift by
\begin{equation}
\delta\nu = \frac{\nu_0}{2} H_0 t,
\end{equation}
which corresponds to a fake acceleration of
\begin{equation}
a_{PIO} = - \frac{H_0 c}{2}.
\end{equation}
This result, however, is still a redshift, instead of the observed blueshift.

\section{Conclusions and outlook}
In this paper we have investigated several ideas on how it might
be possible to explain the much discussed PIONEER anomaly,
which consists of an unexplained blueshift of radiophotons
compared to the predicted values. This effect, which is usually
interpreted in terms of an unexplained acceleration towards the sun
\citep{anderson98,anderson2002}, can not be explained
as due to cosmological acceleration effects. Although
the geodetic motion of an object, like PIONEER-10, in an expanding universe
does lead to residual accelerations in a flat Robertson-Walker-like
universe, these accelerations are off by many orders of magnitude,
and they are also incompatible with the seemingly constant
effect which makes up the anomaly.

We have also demonstrated that, except for a wrong sign,
the order of magnitude of the observed
frequency shift is of the same order as the global cosmological redshift
which occurs when photons propagate freely in the local spacetime.
This redshift has been succesfully applied by countless astronomers
to explain the observed redshift from distant quasar and galaxy
emission lines. The interesting similarity between the
numerical results may also hint at a systematical error
in the physics applied to the analysis of the data.

Since the PIONEER spacecraft is propagating in
a local density fluctuation (the milky way), we have also estimated
the corresponding imprint of the local gravitationally bound system
in a universe with linear expansion. It has been demonstrated that,
although spacetime probably is not expanding in a linear way
(assuming that it is expanding on a local length scale at all),
this effect is, in principle, able to correct the wrong sign
in the local redshift. For these reasons,
more research on this field is strongly encouraged.

Confirming a cosmologically-induced
frequency shift with PIONEER or any upcoming, similarly appropriate,
future spacecraft missions \citep[see, e.g.][]{andturynieto2002}
would help clarify this extremely important problem of
the nature of local spacetime metrics and therefore help to
predict a still ongoing density contrast enhancement on the universe
for time periods in the near future.

\end{document}